\documentclass[11pt]{article}

\usepackage{amsmath, amssymb, amsthm}
\usepackage{geometry}
\usepackage[hidelinks]{hyperref}
\usepackage{tikz}
\usetikzlibrary{arrows.meta,positioning, calc}

\title{\bf Geometric Quantum Computation}
\author{Marco Zaopo\thanks{Email: \texttt{marco.zaopo@gmail.com}}}

\begin{document}
	
	\maketitle
	
	\begin{abstract}
We investigate a geometric route to single-photon quantum-information
encoding, motivated by the Poincarè group extension developed in \cite{ZaopoGroupoid2026}.  In that framework, it is formalized a new class of Lorentz transformations as limits for relative speed going to infinity of ordinary Lorentz boost. Composing pairs of infinite speed boosts obtained on the same direction with limits plus/minus infinite we obtain an involutive real four dimensional matrix
\(C_{\hat n}\), which normalizes the connected Poincar\'e group and induces an index-two extension.  Clifford--Mackey theory then pairs the
future- and past-directed massless Wigner sectors into a single extended
unitary irreducible representation.  The new matrix \(C_{\hat n}\) acts as a unitary sector-exchange
operator \(S_C\),  whose eigenvalues \(\varepsilon=\pm1\) distinguish distinct eigenspaces associated to the extension. We show that its exchange
eigenvectors become entangled relative to a physically selected tensor-product
structure, and give a finite-mode realization in a single photon's
propagation-mode and polarization degrees of freedom.  
Two exchange eigenstates define a single-photon entanglement (SPE) logical
qubit, with coherent controls generating arbitrary single-qubit rotations.
Assuming an ideal nondestructive, coherence-preserving polarization-parity
instrument, we construct logical entangling gates, a controlled-phase Choi
resource, and gate teleportation, establishing conditional circuit-model
universality. 
\end{abstract}
	
	\clearpage
	\tableofcontents
	\clearpage

	\section{Introduction}
	
		Quantum entanglement is usually introduced as a feature of
	composite quantum systems: given a tensor product 
	$\mathcal H_A\otimes\mathcal H_B$, there exist pure states that cannot
	be written as product vectors.  In nonrelativistic quantum mechanics
	this structure is taken as primitive.  By contrast, in relativistic
	quantum theory the state space is constrained by spacetime symmetry:
	one starts from unitary representations of the
	Poincar\'e group and builds quantum fields and particles from them.
	It is then natural to ask whether some instances of entanglement may
	have a direct geometric origin in the representation theory of
	spacetime symmetries.
	
	The mathematical starting point is the index-two extension of the
connected Poincar\'e group generated by the proper non-orthochronous
involution \(C_{\hat n}\) \cite{ZaopoGroupoid2026}.  This matrix can be
obtained as the product of two opposite infinite-speed boost limits along a
fixed spatial direction \(\hat n\).  Only the resulting Poincar\'e extension
is needed here; its construction is recalled briefly in
Section~\ref{subsec:Cn-extended-poincare-group}.

We investigate its massless unitary representations.  The discrete
transformation exchanges the future- and past-directed components of the
null momentum orbit, while the stabilizer of any individual nonzero null
momentum remains \(ISO(2)\).  Clifford--Mackey theory therefore provides a
single extended irreducible representation carrying both sectors.  We ask
when eigenstates of the corresponding exchange operator admit an
operationally meaningful entanglement interpretation and how that structure
can be encoded in single-photon degrees of freedom.

Section~\ref{subsec:Cn-massless-UIRs} derives the massless doublet and its
restriction \eqref{eq:gqc-massless-extension-restriction}.  The corresponding
exchange operator is \eqref{eq:gqc-Cn-exchange-operator}.  Its two eigenvalues
label vectors in a single extended UIR, rather than inequivalent UIRs or an
additional little-group quantum number.

The direct sum alone does not imply entanglement.
Section~\ref{subsec:Cn-zanardi-entanglement} specifies independently accessible
observable algebras and the associated map
\eqref{eq:gqc-operational-sector-isometry}.  Their expectation-value
equivalence is \eqref{eq:gqc-operational-equivalence}; the entanglement and
maximal-entanglement criteria are
\eqref{eq:gqc-entanglement-condition} and
\eqref{eq:gqc-maximal-entanglement-condition}.  Thus the geometric sector
structure and the operational selection of subsystems are distinct inputs.

Section~\ref{sec:experiment} makes the finite-mode realization explicit.
Both laboratory propagation modes have positive energy; they encode selected
abstract state and observable data rather than literally realizing a
past-directed photon four-momentum.  The resulting Bell-like states and
correlation measurements can be implemented within standard single-photon
optics.  Their preparation tests the specified optical encoding, while a
separate empirical test of the spacetime extension would require additional
physical predictions.

We then explore the optical encoding as a computational resource.
Section~\ref{sec:bloch} selects two Bell-like exchange states as a logical
basis and specifies coherent controls generating all of \(SU(2)\) on their
span.  Section~\ref{sec:entangling-two-spe} assumes an additional ideal
nondestructive polarization-parity instrument.  Its restriction to the code
provides logical Pauli-product measurements, from which we construct
entangling gates, a Choi resource and gate teleportation.  Before discussing two-photon operations,
Section~\ref{sec:spe-one-qubit-distinction} makes the key encoding
choice explicit: the correlated single-photon degrees of freedom encode
one SPE logical qubit, rather than two independent logical qubits.
Circuit-model
universality follows under these operational assumptions.

\section{Massless Representations of the \texorpdfstring{\(C_{\hat n}\)}{Cn}-Extended
Poincar\'e Group}
\label{sec:Cn-extended-massless-representations}

\subsection{The \texorpdfstring{\(C_{\hat n}\)}{Cn} Extension of the Poincar\'e Group}
\label{subsec:Cn-extended-poincare-group}

Let \(\eta=\operatorname{diag}(-1,1,1,1)\), and let \(\hat n\in S^2\)
be a unit spatial direction.  The two infinite-speed matrices derived in
\cite{ZaopoGroupoid2026} may be written formally as
\begin{equation*}
 J_{\pm\hat n}=\Lambda_\pm(\hat n)=
 \begin{pmatrix}
  0&\mp\hat n^T\\
  \mp\hat n&I_3-\hat n\hat n^T
 \end{pmatrix},
 \qquad J_{\pm\hat n}^{2}=I.
\end{equation*}
In this paper we need only their product, not the detailed construction
of the limiting transformations.  It gives the real involution
\begin{equation}
 C_{\hat n}:=J_{\hat n}J_{-\hat n}
 =\begin{pmatrix}
   -1&0\\
   0&I_3-2\hat n\hat n^T
  \end{pmatrix}.
 \label{eq:gqc-Cn-matrix}
\end{equation}
In the basis adapted to \(\hat n\), this is
\(\operatorname{diag}(-1,-1,1,1)\).  In particular,
\begin{equation*}
 C_{\hat n}^{2}=I,\qquad
 C_{\hat n}^{T}\eta C_{\hat n}=\eta,\qquad
 \det C_{\hat n}=1,\qquad
 C_{-\hat n}=C_{\hat n}.
\end{equation*}
Thus \(C_{\hat n}\) is proper but non-orthochronous.  It lies outside the
proper orthochronous Lorentz group \(SO^+(\eta)\), which it normalizes.

Let \(\mathbb V\simeq\mathbb R^4\) be the translation group and set
\(\mathcal P_0=\mathbb V\rtimes SO^+(\eta)\).  Conjugation by
\(C_{\hat n}\) induces an involutive automorphism
\begin{equation}
 \alpha_{C_{\hat n}}(a,L)
 =\bigl(C_{\hat n}a,\,C_{\hat n}LC_{\hat n}^{-1}\bigr),
 \qquad \alpha_{C_{\hat n}}^2=\operatorname{id}_{\mathcal P_0}.
 \label{eq:gqc-Cn-poincare-automorphism}
\end{equation}
We henceforth work with the index-two extension
\begin{equation}
 \mathcal P_{\eta,C_{\hat n}}
 =\mathcal P_0\rtimes_{\alpha_{C_{\hat n}}}\mathbb Z_2.
 \label{eq:gqc-Cn-extended-poincare-group}
\end{equation}
For \(g,g'\in\mathcal P_0\) and \(\delta,\delta'\in\{0,1\}\), its product is
\begin{equation*}
 (g,\delta)(g',\delta')
 =\bigl(g\,\alpha_{C_{\hat n}}^\delta(g'),\,\delta+\delta'\bigr),
\end{equation*}
where the second component is evaluated modulo two.  The following
analysis concerns the unitary representations of this extended group,
with emphasis on the action of \(C_{\hat n}\) on massless Wigner sectors.

\subsection{Massless UIRs and the Sector-Exchange Operator}
\label{subsec:Cn-massless-UIRs}

Let
\begin{equation*}
	\pi:
	\mathcal P_0
	\longrightarrow
	U(\mathcal H_\pi)
\end{equation*}
be an ordinary Wigner UIR of the connected Poincaré group.  The automorphism
\(\alpha_{C_{\hat n}}\) defines the conjugate representation
\begin{equation}
	\pi^{C_{\hat n}}(g)
	:=
	\pi\!\left(
	\alpha_{C_{\hat n}}(g)
	\right),
	\qquad
	g\in\mathcal P_0.
	\label{eq:gqc-Cn-conjugate-representation}
\end{equation}

Clifford--Mackey theory distinguishes two possibilities.  If
\begin{equation*}
	\pi^{C_{\hat n}}
	\not\simeq
	\pi,
\end{equation*}
the orbit of \([\pi]\) under the quotient
\(\mathcal P_{\eta,C_{\hat n}}/\mathcal P_0\simeq\mathbb Z_2\)
has two elements.  This is Case A of the Clifford--Mackey alternative, and
the corresponding irreducible representation of the extended group restricts
to
\begin{equation*}
	\operatorname{Res}_{\mathcal P_0}^{\mathcal P_{\eta,C_{\hat n}}}
	\Pi
	\simeq
	\pi\oplus\pi^{C_{\hat n}}.
\end{equation*}
If instead
\begin{equation*}
	\pi^{C_{\hat n}}
	\simeq
	\pi,
\end{equation*}
one obtains Case B, in which \(\pi\) may admit two inequivalent extensions
distinguished by a genuine discrete sign.  The massless construction needed
here belongs to Case A.

Choose the future-directed null representative
\begin{equation}
	k_+
	=
	\omega
	\begin{pmatrix}
		1\\
		\hat n
	\end{pmatrix},
	\qquad
	\omega>0.
	\label{eq:gqc-massless-future-representative}
\end{equation}
The connected point stabilizer is \(ISO(2)\).  No element of the
additional component fixes this future-directed momentum, because that
component reverses time orientation.  The point little group is therefore
not enlarged by the extension.

Its image under the discrete transformation is
\begin{equation}
	k_-
	:=
	C_{\hat n}k_+
	=
	-k_+
	=
	-\omega
	\begin{pmatrix}
		1\\
		\hat n
	\end{pmatrix}.
	\label{eq:gqc-massless-past-representative}
\end{equation}
Thus \(C_{\hat n}\) exchanges the two connected components of the null cone:
\begin{equation*}
	\mathcal O_0^+
	\longleftrightarrow
	\mathcal O_0^-.
\end{equation*}

Let
\begin{equation*}
	\pi_{0,h}^+:
	\mathcal P_0
	\longrightarrow
	U(\mathcal H_{0,h}^+)
\end{equation*}
be a future-directed massless Wigner UIR, where \(h\) denotes the chosen
\(ISO(2)\) little-group representation.  Its \(C_{\hat n}\)-conjugate is a
past-directed massless representation,
\begin{equation*}
	\left(
	\pi_{0,h}^+
	\right)^{C_{\hat n}}
	\simeq
	\pi_{0,h_C}^-,
\end{equation*}
where \(h_C\) denotes the little-group label transformed by conjugation with
\(C_{\hat n}\).

The two representations are inequivalent as representations of
\(\mathcal P_0\), because their translation spectra lie on different
connected components of the null cone:
\begin{equation*}
	\pi_{0,h}^+
	\not\simeq
	\pi_{0,h_C}^-.
\end{equation*}
They therefore form a Case-A doublet.

The corresponding carrier spaces may be written as
\begin{align*}
	\mathcal H_+
	&:=
	\mathcal H_{0,h}^+
	=
	L^2(\mathcal O_0^+,d\mu_0;V_h),
	\\
	\mathcal H_-
	&:=
	\mathcal H_{0,h_C}^-
	=
	L^2(\mathcal O_0^-,d\mu_0;V_{h_C}).
\end{align*}
The map \(k\mapsto C_{\hat n}k\), together with the induced action on the
little-group fiber, determines a unitary transport
\begin{equation*}
	T_0:
	\mathcal H_+
	\longrightarrow
	\mathcal H_-.
\end{equation*}
It satisfies the covariance relation
\begin{equation}
	\pi_{0,h_C}^-(g)
	=
	T_0\,
	\pi_{0,h}^+
	\!\left(
	\alpha_{C_{\hat n}}(g)
	\right)
	T_0^{-1},
	\qquad
	g\in\mathcal P_0.
	\label{eq:gqc-T0-covariance}
\end{equation}
The operator \(T_0\) transports the momentum argument from
\(\mathcal O_0^+\) to \(\mathcal O_0^-\) and simultaneously transports the
corresponding little-group indices.  It is therefore not assumed to be the
identity.

The extended Hilbert space is
\begin{equation}
	\mathcal H_{\mathrm{ext}}
	=
	\mathcal H_+
	\oplus
	\mathcal H_-.
	\label{eq:gqc-extended-massless-Hilbert-space}
\end{equation}
For
\begin{equation*}
	\Psi
	=
	(\psi_+,\psi_-),
	\qquad
	\psi_+\in\mathcal H_+,
	\quad
	\psi_-\in\mathcal H_-,
\end{equation*}
the connected subgroup acts diagonally:
\begin{equation*}
	\Pi(g)
	=
	\begin{pmatrix}
		\pi_{0,h}^+(g)&0\\
		0&\pi_{0,h_C}^-(g)
	\end{pmatrix},
	\qquad
	g\in\mathcal P_0.
\end{equation*}
The discrete generator \(C_{\hat n}\) is represented by
\begin{equation}
	\Pi(C_{\hat n})
	=
	S_C
	:=
	\begin{pmatrix}
		0&T_0^{-1}\\
		T_0&0
	\end{pmatrix}.
	\label{eq:gqc-Cn-exchange-operator}
\end{equation}
Consequently,
\begin{equation*}
	S_C(\psi_+,\psi_-)
	=
	\left(
	T_0^{-1}\psi_-,
	T_0\psi_+
	\right).
\end{equation*}
One immediately obtains
\begin{equation*}
	S_C^2
	=
	I_{\mathcal H_{\mathrm{ext}}}
\end{equation*}
and
\begin{equation}
	S_C\Pi(g)S_C^{-1}
	=
	\Pi\!\left(
	\alpha_{C_{\hat n}}(g)
	\right),
	\qquad
	g\in\mathcal P_0.
	\label{eq:gqc-Cn-semidir-covariance}
\end{equation}
These are precisely the relations required by the semidirect-product
extension.

The resulting extended representation is
\begin{equation}
	\Pi_{0,h}^{\mathrm{ext}}
	=
	\operatorname{Ind}_{\mathcal P_0}^{\mathcal P_{\eta,C_{\hat n}}}
	\pi_{0,h}^+,
	\label{eq:gqc-induced-massless-extension}
\end{equation}
and is irreducible.  Its restriction to the connected subgroup is
\begin{equation}
	\operatorname{Res}_{\mathcal P_0}^{\mathcal P_{\eta,C_{\hat n}}}
	\Pi_{0,h}^{\mathrm{ext}}
	\simeq
	\pi_{0,h}^+
	\oplus
	\pi_{0,h_C}^-.
	\label{eq:gqc-massless-extension-restriction}
\end{equation}

For every normalized \(\psi\in\mathcal H_+\), define
\begin{equation}
	\Psi_\varepsilon(\psi)
	=
	\frac{1}{\sqrt2}
	\left(
	\psi,
	\varepsilon T_0\psi
	\right),
	\qquad
	\varepsilon\in\{+1,-1\}.
	\label{eq:gqc-Cn-exchange-eigenvectors}
\end{equation}
Then
\begin{equation}
	S_C\Psi_\varepsilon(\psi)
	=
	\varepsilon\Psi_\varepsilon(\psi).
	\label{eq:gqc-Cn-exchange-eigenvalue}
\end{equation}

Since the massless doublet belongs to Case A, the two values of
\(\varepsilon\) label eigenvectors of the exchange operator inside the same
extended UIR.  They do not label inequivalent representations of
\(\mathcal P_{\eta,C_{\hat n}}\).  Indeed, replacing \(S_C\) by \(-S_C\)
amounts only to changing the relative phase convention between the two
summands of \(\mathcal H_{\mathrm{ext}}\).

\subsection{Operational Tensor-Product Structure and Entanglement}
\label{subsec:Cn-zanardi-entanglement}

The direct-sum decomposition \eqref{eq:gqc-extended-massless-Hilbert-space}
does not by itself define a bipartite quantum system.  A notion of
entanglement requires the additional selection of a tensor-product structure.

Following Zanardi's operational definition of virtual quantum subsystems
\cite{Zanardi2001Virtual}, the relevant tensor-product structure is selected
by the subalgebras of observables that can be independently accessed,
controlled and measured.  We therefore choose fixed physical
identifications
\begin{equation*}
	R_+:
	\mathcal H_+
	\longrightarrow
	\mathcal H,
	\qquad
	R_-:
	\mathcal H_-
	\longrightarrow
	\mathcal H,
\end{equation*}
and a two-dimensional sector space
\begin{equation*}
	\mathcal K
	=
	\operatorname{span}
	\left\{
	|+\rangle,|-\rangle
	\right\}
	\simeq
	\mathbb C^2.
\end{equation*}
The operational tensor-product identification is the unitary map
\begin{equation*}
	V_{\mathrm{op}}:
	\mathcal H_+\oplus\mathcal H_-
	\longrightarrow
	\mathcal H\otimes\mathcal K
\end{equation*}
defined by
\begin{equation}
	V_{\mathrm{op}}(\psi_+,\psi_-)
	=
	R_+\psi_+\otimes|+\rangle
	+
	R_-\psi_-\otimes|-\rangle.
	\label{eq:gqc-operational-sector-isometry}
\end{equation}

The two operational observable algebras are
\begin{align}
	\mathcal A_{\mathcal H}
	&:=
	V_{\mathrm{op}}^{-1}
	\left(
	\mathcal B(\mathcal H)\otimes I_{\mathcal K}
	\right)
	V_{\mathrm{op}},
	\label{eq:gqc-first-operational-algebra}
	\\
	\mathcal A_{\mathcal K}
	&:=
	V_{\mathrm{op}}^{-1}
	\left(
	I_{\mathcal H}\otimes\mathcal B(\mathcal K)
	\right)
	V_{\mathrm{op}}.
	\label{eq:gqc-second-operational-algebra}
\end{align}
They satisfy
\begin{equation*}
	[
	\mathcal A_{\mathcal H},
	\mathcal A_{\mathcal K}
	]
	=
	0,
\end{equation*}
and together generate the full observable algebra:
\begin{equation*}
	\mathcal A_{\mathcal H}
	\vee
	\mathcal A_{\mathcal K}
	=
	\mathcal B(\mathcal H_{\mathrm{ext}}).
\end{equation*}
Equivalently, each algebra is the commutant of the other.  They therefore
define a bipartite virtual subsystem in the sense of Zanardi.  When the
operations belonging to both algebras are independently realizable in the
laboratory, this tensor-product structure is operationally distinguished
rather than being a purely formal identification.

For
\begin{equation*}
	A\in\mathcal B(\mathcal H),
	\qquad
	B=
	\begin{pmatrix}
		b_{++}&b_{+-}\\
		b_{-+}&b_{--}
	\end{pmatrix}
	\in\mathcal B(\mathcal K),
\end{equation*}
define
\begin{equation}
	\iota_{\mathrm{op}}(A\otimes B)
	:=
	V_{\mathrm{op}}^{-1}
	(A\otimes B)
	V_{\mathrm{op}}.
	\label{eq:gqc-iota-operational-definition}
\end{equation}
In the original direct-sum carrier this is
\begin{equation*}
	\iota_{\mathrm{op}}(A\otimes B)
	=
	\begin{pmatrix}
		b_{++}R_+^{-1}AR_+
		&
		b_{+-}R_+^{-1}AR_-
		\\[2mm]
		b_{-+}R_-^{-1}AR_+
		&
		b_{--}R_-^{-1}AR_-
	\end{pmatrix}.
\end{equation*}
The linear extension of \(\iota_{\mathrm{op}}\) is a unital
\(*\)-isomorphism between
\(\mathcal B(\mathcal H)\otimes\mathcal B(\mathcal K)\)
and
\(\mathcal B(\mathcal H_{\mathrm{ext}})\).

For every
\(\Psi\in\mathcal H_{\mathrm{ext}}\), one therefore has
\begin{equation}
		\left\langle
		\Psi,
		\iota_{\mathrm{op}}(A\otimes B)\Psi
		\right\rangle_{\mathcal H_{\mathrm{ext}}}
		=
		\left\langle
		V_{\mathrm{op}}\Psi,
		(A\otimes B)V_{\mathrm{op}}\Psi
		\right\rangle_{\mathcal H\otimes\mathcal K}.
	\label{eq:gqc-operational-equivalence}
\end{equation}
Equation~\eqref{eq:gqc-operational-equivalence} is the operational
equivalence underlying the virtual-subsystem construction.  It states that
all expectation values of observables belonging to the selected tensor
algebra may be calculated either in the original direct-sum representation
or in the tensor-product representation.

The action of \(T_0\) in the common physical carrier is represented by
\begin{equation}
	\tau_0
	:=
	R_-T_0R_+^{-1}
	\in
	U(\mathcal H).
	\label{eq:gqc-tau0-physical-representative}
\end{equation}
The exchange operator consequently becomes
\begin{equation}
	V_{\mathrm{op}}
	S_C
	V_{\mathrm{op}}^{-1}
	=
	\tau_0^{-1}
	\otimes
	|+\rangle\langle-|
	+
	\tau_0
	\otimes
	|-\rangle\langle+|.
	\label{eq:gqc-exchange-operator-tensor-form}
\end{equation}
If \(\tau_0\) is involutive,
\begin{equation*}
	\tau_0^2=I_{\mathcal H},
\end{equation*}
then
\begin{equation*}
	V_{\mathrm{op}}
	S_C
	V_{\mathrm{op}}^{-1}
	=
	\tau_0\otimes\sigma_x,
\end{equation*}
where
\begin{equation*}
	\sigma_x
	=
	|+\rangle\langle-|
	+
	|-\rangle\langle+|
\end{equation*}
acts on the sector space \(\mathcal K\).

Let
\begin{equation*}
	\phi
	:=
	R_+\psi.
\end{equation*}
Under the operational tensor-product identification, the exchange
eigenvectors in
\eqref{eq:gqc-Cn-exchange-eigenvectors} become
\begin{equation*}
	|\Phi_\varepsilon(\phi)\rangle
	:=
	V_{\mathrm{op}}
	\Psi_\varepsilon(\psi)
	=
	\frac{1}{\sqrt2}
	\left(
	\phi\otimes|+\rangle
	+
	\varepsilon
	\tau_0\phi\otimes|-\rangle
	\right).
\end{equation*}
The reduced state on the sector factor is
\begin{equation*}
	\rho_{\mathcal K}
	=
	\frac12
	\begin{pmatrix}
		1&
		\varepsilon
		\langle\tau_0\phi|\phi\rangle
		\\[2pt]
		\varepsilon
		\langle\phi|\tau_0\phi\rangle
		&
		1
	\end{pmatrix}.
\end{equation*}
Its eigenvalues are
\begin{equation}
	\lambda_\pm
	=
	\frac{
		1
		\pm
		\left|
		\langle\phi|\tau_0\phi\rangle
		\right|
	}{2}.
	\label{eq:gqc-sector-Schmidt-eigenvalues}
\end{equation}
It follows that the exchange eigenstate is entangled relative to the
operational tensor-product structure if and only if
\begin{equation}
	\tau_0\phi
	\not\propto
	\phi.
	\label{eq:gqc-entanglement-condition}
\end{equation}
It is maximally entangled when
\begin{equation}
	\langle\phi|\tau_0\phi\rangle
	=
	0.
	\label{eq:gqc-maximal-entanglement-condition}
\end{equation}

This criterion separates two logically distinct statements.  The
\(C_{\hat n}\)-extended UIR supplies the two-sector carrier, the transport
\(T_0\) and the exchange eigenvectors.  The operationally accessible
observable algebras select the tensor-product structure relative to which
those vectors may be entangled.  Entanglement is therefore not inferred from
the direct-sum decomposition alone.

For the single-mode optical realization, restrict the common carrier to the
two-dimensional propagation-mode space
\begin{equation*}
	\mathcal H_{\mathrm{dir}}
	=
	\operatorname{span}
	\left\{
	|+\hat n\rangle,
	|-\hat n\rangle
	\right\}
	\simeq
	\mathbb C^2
\end{equation*}
and identify the sector space with a two-dimensional polarization space,
\begin{equation*}
	\mathcal K_{\mathrm{pol}}
	=
	\operatorname{span}
	\left\{
	|H\rangle,
	|V\rangle
	\right\}
	\simeq
	\mathbb C^2.
\end{equation*}
Suppose that the physical representative of the transport satisfies
\begin{equation}
	\tau_0|+\hat n\rangle
	=
	e^{i\chi}|-\hat n\rangle,
	\label{eq:gqc-optical-T0-action}
\end{equation}
where the phase \(\chi\) can be absorbed into the definition of one of the
sector basis vectors.  Then
\begin{equation*}
	\langle+\hat n|
	\tau_0
	|+\hat n\rangle
	=
	0,
\end{equation*}
and the exchange eigenvectors become the Bell-like single-photon states
\begin{equation}
	|\Phi_\varepsilon\rangle
	=
	\frac{1}{\sqrt2}
	\left(
	|+\hat n\rangle\otimes|H\rangle
	+
	\varepsilon
	|-\hat n\rangle\otimes|V\rangle
	\right).
	\label{eq:gqc-single-photon-Bell-like-states}
\end{equation}
With the phase convention in which
\begin{equation*}
	\tau_0
	=
	\sigma_x^{(\mathrm{dir})},
\end{equation*}
one has
\begin{equation}
	V_{\mathrm{op}}
	S_C
	V_{\mathrm{op}}^{-1}
	=
	\sigma_x^{(\mathrm{dir})}
	\otimes
	\sigma_x^{(\mathrm{pol})}.
	\label{eq:gqc-optical-exchange-operator}
\end{equation}
The states in
\eqref{eq:gqc-single-photon-Bell-like-states}
are its eigenstates with eigenvalues
\(\varepsilon=\pm1\).

A different identification of the two carrier spaces may absorb \(T_0\) and
make \(\tau_0=I\).  Relative to that alternative tensor-product structure,
the same exchange eigenvectors are product states.  This is not a
contradiction: it is precisely the relativity of entanglement with respect to
the selected tensor-product structure emphasized by Zanardi.  The physical
claim of the present construction is that the independently controllable
propagation-mode and polarization observables select
\(\mathcal H_{\mathrm{dir}}\otimes\mathcal K_{\mathrm{pol}}\)
as the operationally relevant tensor-product structure.

The representation-theoretic sectors
\(\mathcal H_+\) and \(\mathcal H_-\) refer, at the abstract Wigner level, to
the future- and past-directed components of the null cone.  Their realization
as experimentally accessible positive-frequency counterpropagating photon
modes requires an explicit physical realization map.  This realization is
an additional step, developed in the following section, and should not be
identified with the abstract Clifford--Mackey classification itself.

The logical chain combines the doublet
\eqref{eq:gqc-extended-massless-Hilbert-space}, the operational algebras
\eqref{eq:gqc-first-operational-algebra}--\eqref{eq:gqc-second-operational-algebra},
and the optical transport condition \eqref{eq:gqc-optical-T0-action} to obtain
\eqref{eq:gqc-single-photon-Bell-like-states}.

In this precise sense, the \(C_{\hat n}\)-extended massless UIR supplies the
representation-theoretic structure from which an operationally meaningful
single-photon entangled encoding can be constructed.  The entanglement does
not follow from a Hilbert-space isomorphism alone; it follows from the
extended-sector transport together with the physically selected and
experimentally accessible tensor-product structure.

	%---------------------------------------------------------
	% EXPERIMENTAL PROPOSAL
	%---------------------------------------------------------
	
	\section{Experimental Access to the Massless Sector-Exchange Structure}
\label{sec:experiment}

We now implement the selected exchange states and observable algebra of
Section~\ref{subsec:Cn-zanardi-entanglement} with a single photon.  The binary
value \(\varepsilon\) is the eigenvalue of \(S_C\); it does not distinguish
two extended representations.

\subsection{Fixed Axis and Finite-Mode Optical Realization}
\label{subsec:optical-realization}

Write the propagation direction as
\begin{equation}
 \hat n=(\sin\theta\cos\phi,\sin\theta\sin\phi,\cos\theta).
 \label{eq:optical-axis}
\end{equation}
The involution \eqref{eq:gqc-Cn-matrix} depends only on the unoriented axis
\([\hat n]\).  We choose \(\hat n=\hat z\) for the laboratory setup.
The abstract representatives \(k_+\) and \(k_-\) are those of
\eqref{eq:gqc-massless-future-representative} and
\eqref{eq:gqc-massless-past-representative}; their zeroth components have
opposite signs.  The two physical optical modes instead have momenta
\begin{equation}
 p^{\mathrm{opt}}_\delta=\omega(1,\delta\hat n),
 \qquad \omega>0,\quad \delta=\pm1.
 \label{eq:positive-frequency-optical-momenta}
\end{equation}
In particular, \(p^{\mathrm{opt}}_-\neq k_-\).  The optical mode label
\(\delta\), the abstract energy-branch label, and the exchange eigenvalue
\(\varepsilon\) must be distinguished.

To specify the realization, select orthonormal wave packets
\(f_+,f_-\in\mathcal H_+\) and their images under \(T_0\).  On the
four-dimensional subspace
\(\mathcal F=\operatorname{span}\{f_+,f_-\}\oplus
 T_0\operatorname{span}\{f_+,f_-\}\), define
\begin{equation}
 \mathcal R_{\mathrm{opt}}(f_\delta,0)=|\delta\rangle|H\rangle,
 \qquad
 \mathcal R_{\mathrm{opt}}(0,T_0f_\delta)=|-\delta\rangle|V\rangle.
 \label{eq:finite-mode-realization-map}
\end{equation}
Here \(|\delta\rangle\equiv|\delta\hat n\rangle\) denotes the laboratory
mode, not an abstract energy sign.  The orthonormal bases make
\(\mathcal R_{\mathrm{opt}}\) unitary onto
\(\mathcal H_{\mathrm{dir}}\otimes\mathcal K_{\mathrm{pol}}\).
This is the finite-mode instance of \(V_{\mathrm{op}}\) with
\(R_+f_\delta=|\delta\rangle\) and
\(R_-T_0f_\delta=|-\delta\rangle\).  It gives exactly the exchange
operator \eqref{eq:gqc-optical-exchange-operator} and maps
\(\Psi_\varepsilon(f_+)\) to the optical state
\eqref{eq:gqc-single-photon-Bell-like-states}.

The selected subspace is invariant under \(S_C\), but need not be invariant
under the full Poincar\'e action.  This map realizes the chosen exchange and
observable structure; it is not asserted to intertwine the full abstract
translation representation with the positive-frequency photon
representation.  Both polarizations are used in this encoding, so it is
not a restriction to a single physical photon helicity.

A tunable relative phase prepares the family
\begin{equation}
 |\Psi(\varphi)\rangle=
 \frac{|+\rangle|H\rangle+e^{i\varphi}|-\rangle|V\rangle}{\sqrt2}.
 \label{eq:Psi-varphi}
\end{equation}
Its members at \(\varphi=0\) and \(\varphi=\pi\), modulo \(2\pi\),
are respectively \(|\Phi_+\rangle\) and \(|\Phi_-\rangle\) of
\eqref{eq:gqc-single-photon-Bell-like-states}.  At other phases the state
need not be an exchange eigenstate.

\subsection{Preparation Stage}

A preparation scheme for \eqref{eq:Psi-varphi} is:
\begin{enumerate}
 \item Prepare a heralded single-photon wave packet in a known polarization
 and split it coherently between two balanced optical modes.
 \item Route the modes to encode propagation along \(\pm\hat n\), using
 mirrors and mode matching as needed.  Direction is the common-carrier
 factor in \eqref{eq:finite-mode-realization-map}.
 \item Use polarization optics to associate the two modes with \(H\) and
 \(V\), and control their relative phase.  The two phase choices above
 prepare the opposite exchange eigenvalues within the same code.
\end{enumerate}
The chosen optical identification is an operational assumption; splitting
or reflecting a positive-frequency photon does not change its energy sign.

\subsection{Measurement Stage and Dependence on \(\varepsilon\)}

Denote the optical representative in \eqref{eq:gqc-optical-exchange-operator}
by \(\overline S_C\), and write \(O_{XX}:=\overline S_C\).
Its abstract observable is obtained through
\eqref{eq:gqc-iota-operational-definition}.  The direction Pauli operator
mixes the two optical modes, while the polarization Pauli operator mixes
\(H\) and \(V\).  Direct evaluation on \eqref{eq:Psi-varphi} gives
\begin{equation}
 \langle\Psi(\varphi)|O_{XX}|\Psi(\varphi)\rangle=\cos\varphi,
 \qquad
 \langle\Phi_\varepsilon|O_{XX}|\Phi_\varepsilon\rangle=\varepsilon.
 \label{eq:XX-expectation}
\end{equation}
The two optical-factor marginals are both \(I_2/2\), independently of the
relative phase.  The exchange sign belongs to their joint correlation,
not to either individual polarization or direction qubit.

Route and mode-match the two propagation modes at a balanced beam splitter
to measure their \(X\) basis.  In each output port, a polarization basis
rotation followed by a polarizing beam splitter measures the polarization
\(X\) basis.  Single-photon detection yields joint outcomes
\((x_{\mathrm{dir}},x_{\mathrm{pol}})\in\{\pm1\}^2\), from which
\begin{equation*}
 E_{XX}=\sum_{x_{\mathrm{dir}},x_{\mathrm{pol}}}
 x_{\mathrm{dir}}x_{\mathrm{pol}}
 P(x_{\mathrm{dir}},x_{\mathrm{pol}})
\end{equation*}
is reconstructed.  This correlation distinguishes the selected eigenstates;
full state tomography additionally requires other measurement settings.

\subsection{Interpretation and Scope of the Test}

The phase-dependent correlation \eqref{eq:XX-expectation} is also predicted
by ordinary quantum optics.  Measuring it verifies the prepared optical
state and the chosen realization of the exchange observable.  It does not,
by itself, discriminate the spacetime extension from standard quantum
mechanics.  Conversely, failure to prepare one of the signs would require
checking optical control and calibration before drawing conclusions about
the proposed interpretation.

The additional geometric claim is the correspondence between this accessible
optical algebra and the abstract sector-exchange construction.  A distinct
falsifiable prediction for that claim requires further assumptions connecting
the full spacetime action or dynamics to physical observables.  No such
discriminating prediction follows solely from the finite-mode map
\eqref{eq:finite-mode-realization-map}.

\subsection{Relation to Existing Single--Particle Entanglement Experiments}
	\label{subsec:existing-experiments}
	
	The state and measurement we propose in this section are,
	from the point of view of standard quantum optics, instances of
	single--particle entanglement between two internal degrees of freedom.
	In particular, the Bell-like state
\eqref{eq:gqc-single-photon-Bell-like-states}
is a special case of the generic path--polarisation entangled state of a
	single photon, where \(|+\rangle,|-\rangle\) encode two spatial modes and
	\(|H\rangle,|V\rangle\) encode two polarisation modes.  Correlations of
	Pauli--type observables such as
	\(\sigma_x^{(\mathrm{dir})}\otimes\sigma_x^{(\mathrm{pol})}\) have been
	measured in several experiments.
	
	For example, Fiorentino \emph{et al.} implement a deterministic
	controlled--NOT gate acting on two qubits carried by a \emph{single}
	photon: one qubit is encoded in polarisation, the other in the spatial
	mode (momentum) of the photon \cite{Fiorentino2004}.  Their preparation and analysis stages access path--polarization states
of the form \eqref{eq:Psi-varphi}, after relabelling the two spatial modes.
The phase choices \(0,\pi\) correspond to
\eqref{eq:gqc-single-photon-Bell-like-states}.

Full two--qubit tomography in mutually
	unbiased bases (including the \(X\) basis on each qubit) is performed,
	so that correlations proportional to
	\(\langle\sigma_x^{(\mathrm{path})}\otimes\sigma_x^{(\mathrm{pol})}\rangle\)
	are effectively accessed \cite{Fiorentino2004}.
	
	Similarly, Bera \emph{et al.} propose a protocol in which intra--photon
	entanglement between path and polarisation is swapped to inter--photon
	entanglement using linear optics \cite{Bera2018}.  Their initial
	single--photon resource is again a maximally entangled state between
	path and polarisation, structurally identical to
	\(|\Phi_\varepsilon\rangle\), and the protocol requires the ability to
	prepare, transform and analyse such states in various local bases.
	
	Beyond photonic systems, Hasegawa \emph{et al.} demonstrated
	entanglement between path and spin degrees of freedom of single neutrons
	in a Mach–Zehnder–type interferometer \cite{Hasegawa2003}.  There, the
	two--dimensional ``path'' subspace and the spin--\(\tfrac12\) space play
	the role of the two qubits, and joint spin--path measurements are
	performed in different bases to reveal nonclassical correlations.
	
	From the purely operational point of view, these experiments show that
	
	\begin{itemize}
		\item a state preparation equivalent (up to local unitaries) to
		\(|\Phi_\varepsilon\rangle\) is experimentally standard,
		\item joint measurements of Pauli observables on the two internal
		degrees of freedom---including \(X\otimes X\)--type correlations---are
		feasible and have been performed, and
		\item the observed correlations agree with the quantum predictions for
		a maximally entangled two--qubit state within experimental
		accuracy.
	\end{itemize}
	
	The novelty of the present work does not lie in the optical technology
	required to prepare and measure such single--photon two--qubit states,
	but in the \emph{representation--theoretic interpretation}.  In the
	above experiments, the two qubits are treated as abstract, independent
	Hilbert--space degrees of freedom (path, polarisation, spin), and their
	entanglement is a kinematical feature of the chosen encoding.  In the present framework, the independently controlled optical factors
select the tensor-product structure of
Section~\ref{subsec:Cn-zanardi-entanglement}.  Their relation to the abstract
energy branches is the additional realization map
\eqref{eq:finite-mode-realization-map}; neither optical factor alone is the
exchange eigenvalue \(\varepsilon\).

Existing experiments access the same state and correlation structure.
Reinterpreting these data through \(C_{\hat n}\) does not constitute an
independent verification of the extended spacetime symmetry.  It supplies
an operational setting for the encoding and computation developed below.

%=========================================================

\section{Single Photon Entanglement Qubit}
	\label{sec:bloch}
	
	In this section we show that entangled two qubits states introduced in the previous section, together with a tunable relative phase shifter and a two modes photons coupler, can be used to operatively construct the manifold of possible states of a single qubit, i.e. a Bloch sphere. We will call the corresponding idealized system single photon entanglement qubit. In section \ref{subsec:encoding} we will encode perfectly distinguishable states in two selected eigenstates of $\overline S_C$, while in section \ref{subsec:blochsphere} we will show that the chosen encoding allows access of the whole Bloch sphere by means of a tunable relative phase shifter and a photonic two modes coupler.  
	
	\subsection{Encoding}
	\label{subsec:encoding}
	
	Fix the laboratory axis \eqref{eq:optical-axis} and the realization map
\eqref{eq:finite-mode-realization-map}.  The optical exchange operator
\(\overline S_C\) of \eqref{eq:gqc-optical-exchange-operator} has two
\emph{two-dimensional} eigenspaces on the full four-dimensional optical
carrier.  We select the pair \(|\Phi_+\rangle,|\Phi_-\rangle\) in
\eqref{eq:gqc-single-photon-Bell-like-states} to define one logical qubit;
the axis or exchange spectrum alone does not select this code.
For compact notation, define

\[
	|\Phi_{+H}\rangle := |+\rangle\otimes|H\rangle,\qquad
	|\Phi_{-V}\rangle := |-\rangle\otimes|V\rangle,
	\]
	where $|+\rangle,|-\rangle$ label propagation direction along $\pm\hat n$
	and $|H\rangle,|V\rangle$ label two orthogonal internal modes (e.g. polarisation).
	The relevant two-mode physical subspace is
	
	\begin{equation}\label{H2m}
		\mathcal H_{\mathrm{2m}}:=\mathrm{span}\{|\Phi_{+H}\rangle,|\Phi_{-V}\rangle\}\subset \mathbb C^2_{\rm dir}\otimes\mathbb C^2_{\rm pol}.
	\end{equation}
	
	The selected exchange states define
\begin{equation}
 |0_L\rangle:=|\Phi_+\rangle,\qquad
 |1_L\rangle:=|\Phi_-\rangle,\qquad
 \mathcal C:=\operatorname{span}\{|0_L\rangle,|1_L\rangle\}
 =\mathcal H_{\mathrm{2m}}.
 \label{eq:logical-code}
\end{equation}
Here \(\overline S_C|_{\mathcal C}=Z_L\).  A general logical state need not
be an exchange eigenstate or be entangled between the two physical factors;
for example, a balanced sum of the logical basis states is
\(|\Phi_{+H}\rangle\).  The name SPE refers to the entangled logical basis,
not to a requirement that every vector of the code be entangled.

Starting from the logical states defined above $|0_L\rangle$ and  $|1_L\rangle$ one can define a generic state of the Bloch sphere as:
	\begin{equation}
		|\psi_L(\vartheta,\varphi)\rangle
		=
		\cos\frac{\vartheta}{2}\,|0_L\rangle
		+
		e^{i\varphi}\sin\frac{\vartheta}{2}\,|1_L\rangle,
		\qquad
		0\le\vartheta\le\pi,\ \ 0\le\varphi<2\pi.
		\label{eq:logical-state-theta-phi}
	\end{equation}
	In the physical two-mode embedding $\mathcal H_{\mathrm{2m}}=\mathrm{span}\{|\Phi_{+H}\rangle,|\Phi_{-V}\rangle\}$ this becomes
	\begin{equation*}
		|\psi_L(\vartheta,\varphi)\rangle
		=
		\frac{1}{\sqrt2}\Big(
		(\cos\frac{\vartheta}{2}+e^{i\varphi}\sin\frac{\vartheta}{2})\,|\Phi_{+H}\rangle
		+
		(\cos\frac{\vartheta}{2}-e^{i\varphi}\sin\frac{\vartheta}{2})\,|\Phi_{-V}\rangle
		\Big),
	\end{equation*}

	\subsection{Physical Realization of the Bloch Sphere}
	\label{subsec:blochsphere}
	
	We are now going to show that any state of the Bloch sphere defined above can be reached exploiting a relative phase shifter and a two mode coupler acting on the space defined in \eqref{H2m}. 
	
	Combining \eqref{eq:logical-code} with
\eqref{eq:gqc-single-photon-Bell-like-states} gives
	\begin{equation*}
		\begin{pmatrix}|\Phi_{+H}\rangle\\|\Phi_{-V}\rangle\end{pmatrix}
		=
		H_L
		\begin{pmatrix}|0_L\rangle\\|1_L\rangle\end{pmatrix},
		\qquad
		H_L=\frac{1}{\sqrt2}\begin{pmatrix}1&1\\[2pt]1&-1\end{pmatrix},
	\end{equation*}
	so that the matrix of an operator $U$ in the logical basis is
	\begin{equation}
		U_L = H_L^\dagger U H_L.
		\label{eq:conjugation_to_logical}
	\end{equation}
	Since $H_L^\dagger=H_L=H_L^{-1}$, this is just a Hadamard conjugation.

	We assume two coherent optical controls that preserve the code.  Because
its two basis modes have orthogonal polarizations, a polarization-insensitive
beam splitter alone does not mix them within the code: suitable mode routing
and polarization conversion must implement the effective coupler below.
Both controls act on
	$\mathcal H_{\mathrm{2m}}$:
	\begin{enumerate}
		\item A relative phase shifter
		\begin{equation}
			P(\alpha)=\mathrm{diag}(e^{-i\alpha/2},\,e^{i\alpha/2})
			=
			\exp\!\Big(-\frac{i}{2}\alpha\,\sigma_z\Big)
			\quad\text{in the basis }\{|\Phi_{+H}\rangle,|\Phi_{-V}\rangle\},
			\label{eq:phase_shifter_ab}
		\end{equation}
		so its generator is $-\frac{i}{2}\sigma_z$.
		
		\item A coherent coupler mixing the two selected composite modes,
		\begin{equation}
			B(\beta)=
			\begin{pmatrix}
				\cos(\beta/2) & -\sin(\beta/2)\\
				\sin(\beta/2) & \ \cos(\beta/2)
			\end{pmatrix}
			=
			\exp\!\Big(-\frac{i}{2}\beta\,\sigma_y\Big)
			\quad\text{in the basis }\{|\Phi_{+H}\rangle,|\Phi_{-V}\rangle\},
			\label{eq:coupler_ab}
		\end{equation}
		so its generator is $-\frac{i}{2}\sigma_y$.
	\end{enumerate}
	
	Using \eqref{eq:conjugation_to_logical} and
\(H_L\sigma_zH_L=\sigma_x\), \(H_L\sigma_yH_L=-\sigma_y\), one obtains
\begin{equation}
		P_L(\alpha)=\exp\!\Big(-\frac{i}{2}\alpha\,\sigma_x\Big),
		\qquad
		B_L(\beta)=\exp\!\Big(-\frac{i}{2}\beta\,(-\sigma_y)\Big)
		=\exp\!\Big(+\frac{i}{2}\beta\,\sigma_y\Big).
		\label{eq:PLBL_logical_paulis}
	\end{equation}
	Define \(X_L,Y_L,Z_L\) as the standard Pauli matrices in the ordered
basis \((|0_L\rangle,|1_L\rangle)\).  With \(\widetilde B_L(\beta):=B_L(-\beta)\),
the controls in \eqref{eq:PLBL_logical_paulis} generate rotations about
\(X_L\) and \(Y_L\).  The reparametrization records the sign change explicitly.

Consider the Lie algebra generated by $\{iX_L,iY_L\}$.
	Using the Pauli commutation relations,
	\begin{equation*}
		[X_L,Y_L]=2iZ_L,
	\end{equation*}
	we see that $iZ_L$ belongs to the generated Lie algebra as well. Since
	$\{iX_L,iY_L,iZ_L\}$ is a basis of $\mathfrak{su}(2)$, it follows that
	\begin{equation}
		\mathrm{Lie}\langle iX_L,iY_L\rangle=\mathfrak{su}(2).
		\label{eq:lie_generated_su2}
	\end{equation}

	Within the logical unitary group $SU(2)$, the
	connected subgroup generated by the two tunable one--parameter families
	$\{P_L(\alpha)\}_{\alpha\in\mathbb R}$ and $\{B_L(\beta)\}_{\beta\in\mathbb R}$
	is all of $SU(2)$ acting on $\mathcal C$. 
	This proves that the SPE code supports arbitrary logical single--qubit unitaries
	using only the physically available phase shifter and two--mode coupler, once
	their action is represented in the $\overline S_C$ eigenbasis of the code.
%=========================================================
\section{Single-Photon Entanglement as a One-Qubit Encoding}
\label{sec:spe-one-qubit-distinction}

The computational use proposed here differs from the familiar treatment of
path and polarization as two separately addressable logical qubits carried
by the same photon, as in the single-photon two-qubit experiments discussed
in Section~\ref{subsec:existing-experiments} \cite{Fiorentino2004}.
More generally, multi-degree-of-freedom and hyperentangled photonic schemes
may use distinct degrees of freedom as separate logical subsystems or as
resources for multi-qubit protocols.  Our proposal instead uses the
\emph{entanglement between two degrees of freedom of one photon to encode
one logical qubit}.  The two physical factors provide the correlations
that carry the logical information; they are not interpreted as two
independently encoded computational qubits within this code.

Specifically, the four-dimensional physical carrier
$\mathcal H_{\rm dir}\otimes\mathcal H_{\rm pol}$ contains the selected
\emph{two-dimensional} code $\mathcal C$ of
Eq.~\eqref{eq:logical-code}.  Its computational basis is
\begin{equation}
 \begin{aligned}
  |0_L\rangle &= |\Phi_+\rangle
  =\frac{|+\rangle|H\rangle+|-\rangle|V\rangle}{\sqrt2},\\
  |1_L\rangle &= |\Phi_-\rangle
  =\frac{|+\rangle|H\rangle-|-\rangle|V\rangle}{\sqrt2}.
 \end{aligned}
 \label{eq:spe-correlation-logical-basis}
\end{equation}
Both basis states have the same maximally mixed marginals on propagation
mode and polarization.  A measurement of either degree of freedom alone
therefore reveals no logical bit.  Likewise, both states are perfectly
correlated in the $Z\otimes Z$ basis, since
\begin{equation*}
 (Z_{\rm dir}\otimes Z_{\rm pol})|0_L\rangle=|0_L\rangle,
 \qquad
 (Z_{\rm dir}\otimes Z_{\rm pol})|1_L\rangle=|1_L\rangle.
\end{equation*}
The logical values are instead distinguished by the \emph{sign of their
joint $X$-basis correlation}, which is the exchange observable of the
selected optical realization:
\begin{equation}
 (X_{\rm dir}\otimes X_{\rm pol})|0_L\rangle=+|0_L\rangle,
 \qquad
 (X_{\rm dir}\otimes X_{\rm pol})|1_L\rangle=-|1_L\rangle,
 \qquad
 Z_L=\overline S_C\big|_{\mathcal C}.
 \label{eq:spe-correlation-encodes-logical-Z}
\end{equation}
If $x_{\rm dir},x_{\rm pol}\in\{+1,-1\}$ are the outcomes of the two
$X$-basis measurements on that \emph{single} photon, their joint
probabilities are
\begin{equation}
 \Pr(x_{\rm dir},x_{\rm pol}\mid |0_L\rangle)
 =\frac{1+x_{\rm dir}x_{\rm pol}}{4},\qquad
 \Pr(x_{\rm dir},x_{\rm pol}\mid |1_L\rangle)
 =\frac{1-x_{\rm dir}x_{\rm pol}}{4}.
 \label{eq:spe-correlated-outcome-probabilities}
\end{equation}
Thus $|0_L\rangle$ gives perfectly \emph{correlated} outcomes,
$x_{\rm dir}x_{\rm pol}=+1$, whereas $|1_L\rangle$ gives perfectly
\emph{anticorrelated} outcomes, $x_{\rm dir}x_{\rm pol}=-1$.
The logical outcome is the product of the two physical outcomes, not either
outcome taken separately.  This is a single-photon joint-correlation
readout of one logical qubit.

The encoding is not restricted to classical logical bits.  Arbitrary
coherent superpositions of these two states form the Bloch sphere
\eqref{eq:logical-state-theta-phi}, with unitary controls constructed in
Section~\ref{subsec:blochsphere}.  Such superpositions need not themselves
be maximally entangled, or even entangled, between the physical factors.
What defines the SPE encoding is the chosen \emph{entangled logical basis}
and its operational correlation observable, not an assertion that every
logical state is entangled.

Consequently, \emph{one photon carries one SPE logical qubit} in the
architecture studied here.  Two photons encode two logical SPE qubits,
each defined by its own internal path--polarization correlations.  The
inter-photon operations of the following section entangle these
\emph{logical qubits}; they must not be confused with the intra-photon
entanglement already used to define each qubit.  The use of a
four-dimensional physical carrier for a two-dimensional logical code
alone does not imply fault tolerance or a scalability advantage; those
questions require an implementation and a noise model.

	%=========================================================
	\section{Entangling Two Single Photon Entanglement Qubits}
	\label{sec:entangling-two-spe}
	
	In this section we construct a logical entangling gate using a coherent single--qubit $SU(2)$ control on the code space $\mathcal C$ (see section ~\ref{sec:bloch}) and
	a two--photon polarisation parity measurement, which is a two--outcome CPTP instrument on the underlying optical degrees of freedom.
	
	In order to establish the result we will proceed as follows:

	\begin{enumerate}
		\item We write the parity measurement as a quantum instrument defined on two SPE qubits state space in Kraus form \cite{Krauss} and explicit it in the logical basis.
		\item We show that the the logical Pauli products defined in \eqref{eq:S1S2_def}
have a unique common $+1$ eigenstate and is maximally entangled.
		\item We derive the Choi--Jamio\l kowski \cite{Choi}, \cite{Jamiol} resource state associated with the logical $CZ_L$ channel
		and show it is the unique common $+1$ eigenstate of the commuting stabilisers $\widetilde S_i$, $i=1,2,3,4$ representing parity polarization measurements on four SPE qubits.
		\item We obtain a logical $CNOT_L$ gate up to Pauli products in two SPE qubits logical state space via a teleportation protocol built from the Choi representation defined at previous point.
	\end{enumerate}
	
	%---------------------------------------------------------
	\subsection{Parity Polarisation Measurement as a CPTP Instrument}
	\label{subsec:parity-instrument}
	
	On two photons $A,B$ we consider the physical polarisation parity observable
	\begin{equation*}
		M_{ZZ}^{(\mathrm{pol})}
		:=
		I_{\rm dir}^{(A)}\otimes \sigma_z^{(\mathrm{pol},A)}
		\otimes
		I_{\rm dir}^{(B)}\otimes \sigma_z^{(\mathrm{pol},B)},
		\qquad
		\sigma_z|H\rangle=|H\rangle,\ \sigma_z|V\rangle=-|V\rangle.
	\end{equation*}
	The two outcomes $s=\pm 1$ are described by projectors
	\begin{equation*}
		\Pi_s^{(\mathrm{pol})}
		=
		\frac12\Big(I + s\,\sigma_z^{(\mathrm{pol},A)}\sigma_z^{(\mathrm{pol},B)}\Big),
	\end{equation*}
	and ideal parity readout is the two--outcome instrument
	\begin{equation}
		\mathcal E_s(\rho)=\Pi_s^{(\mathrm{pol})}\,\rho\,\Pi_s^{(\mathrm{pol})},
		\qquad s\in\{+1,-1\}.
		\label{eq:instrument_pol}
	\end{equation}
	Thus, conditioned on $s$, the map has a single Kraus operator $K_s=\Pi_s^{(\mathrm{pol})}$; and
	the unconditional map $\mathcal E=\mathcal E_{+}+\mathcal E_{-}$ is CPTP.
We assume an ideal nondestructive implementation that retains both photons
and preserves coherence within each parity eigenspace.  Measuring the two
polarizations separately and multiplying their outcomes is not this
instrument.  Its physical realization and success probability are resource
assumptions of the computational model, not consequences of the group
extension or of single-photon linear optics alone.

	Use the physical modes in \eqref{H2m} and the logical basis
\eqref{eq:logical-code}.

Since
	\(
	\sigma_z^{(\mathrm{pol})}|\Phi_{+H}\rangle=+|\Phi_{+H}\rangle
	\)
	and
	\(
	\sigma_z^{(\mathrm{pol})}|\Phi_{-V}\rangle=-|\Phi_{-V}\rangle,
	\)
	one finds
	\begin{equation*}
		\sigma_z^{(\mathrm{pol})}|0_L\rangle=|1_L\rangle,
		\qquad
		\sigma_z^{(\mathrm{pol})}|1_L\rangle=|0_L\rangle,
		\qquad\Rightarrow\qquad
		\sigma_z^{(\mathrm{pol})}\big|_{\mathcal C}=X_L.
	\end{equation*}
	Therefore, on $\mathcal C_A\otimes\mathcal C_B$,
	\begin{equation}
		\Big(\sigma_z^{(\mathrm{pol},A)}\sigma_z^{(\mathrm{pol},B)}\Big)\Big|_{\mathcal C^{(2)}}
		=
		X_L^{(A)}X_L^{(B)},
		\qquad
		\Pi_s^{(\mathrm{pol})}\Big|_{\mathcal C^{(2)}}
		=
		\frac12\Big(I + s\,X_L^{(A)}X_L^{(B)}\Big).
		\label{eq:parity_restricts_to_XX}
	\end{equation}
	Hence, in the ideal code-preserving limit, the physical parity device realises a
	logical Pauli--product measurement of $X_L\otimes X_L$. Using available single--qubit $SU(2)$
	rotations, the same primitive realises any Pauli--product measurement needed below (e.g.\ $Z_LZ_L$).
	
	%---------------------------------------------------------
	\subsection{Existence of Entangled States From Parity Polarization Measurements}
	\label{subsec:C2_from_S1S2}
	
	Consider two logical SPE qubits labelled $(C,D)$. Define the commuting Pauli products
	\begin{equation}
		S_1:=X_L^{(C)}Z_L^{(D)},\qquad S_2:=Z_L^{(C)}X_L^{(D)}.
		\label{eq:S1S2_def}
	\end{equation}
	Commutation can be verified by the reader.
	
	We now determine the unique common $+1$ eigenstate explicitly.
	Write a general logical two--qubit state as
	\[
	|\psi\rangle
	=
	a|0_L0_L\rangle+b|0_L1_L\rangle+c|1_L0_L\rangle+d|1_L1_L\rangle.
	\]
	Imposing $S_1|\psi\rangle=|\psi\rangle$ and $S_2|\psi\rangle=|\psi\rangle$ gives
	\[
	c=a,\qquad d=-b,\qquad b=a,\qquad d=-c,
	\]
	hence $b=c=a$ and $d=-a$. Up to normalisation, there is a unique solution:
	\begin{equation}
		|C_2\rangle_{CD}
		=
		\frac12\Big(|0_L0_L\rangle+|0_L1_L\rangle+|1_L0_L\rangle-|1_L1_L\rangle\Big)_{CD}.
		\label{eq:C2_explicit}
	\end{equation}
	This state is maximally entangled. Indeed, the coefficient matrix in the logical basis is
	\(
	M=\frac12\begin{pmatrix}1&1\\ 1&-1\end{pmatrix}
	\),
	so $MM^\dagger=\mathbf 1/2$ and the reduced state is maximally mixed.
	
	Operationally, $|C_2\rangle$ is prepared inside $\mathcal C^{(2)}$ by measuring $S_1$ and $S_2$.
	If one obtains the $(-1)$ outcome for either stabiliser, the resulting state differs from
	\eqref{eq:C2_explicit} by a \emph{known} local Pauli byproduct (tracked classically or corrected).
	
	%---------------------------------------------------------
	\subsection{Controlled-$NOT$ Gate Construction}
	\label{subsec:RCZ_stabiliser}
	
	In this subsection we consider the above defined parity polarization measurements on four SPE qubits, represent it in logical basis and show it has a unique eigenstate with eigenvalue +1. We call such state $|R_{CZ}\rangle$. We then show that  $|R_{CZ}\rangle$ can be used to construct a $CNOT_L$ gate defined on two logical qubits up to Pauli products exploiting a suitable teleportation protocol.

	Introduce four logical qubits $C,C',D,D'$ and consider the following set of measurements:
	\begin{equation}
		\begin{aligned}
			\widetilde S_1 &:= Z_L^{(C)}Z_L^{(C')},\\
			\widetilde S_2 &:= Z_L^{(D)}Z_L^{(D')},\\
			\widetilde S_3 &:= X_L^{(C)}X_L^{(C')}Z_L^{(D)},\\
			\widetilde S_4 &:= X_L^{(D)}X_L^{(D')}Z_L^{(C)}.
		\end{aligned}
		\label{eq:RCZ_stabilizers}
	\end{equation}
	
	In the next subsection we are going to construct explicitly these three qubits measurements via ancilla assisted reduction of three qubits Pauli tensor products measurements to two qubits polarization parity measurements plus ancillas.
	
	%---------------------------------------------------------
	\subsubsection{From Three Qubits Measurements to Polarization Parity Measurements}
	\label{subsubsec:3body_to_2body}
	
	The three-body stabilizers in \eqref{eq:RCZ_stabilizers} require a
measurement of their joint parity that preserves coherence within each
eigenspace.  Measuring several commuting pairwise parities and multiplying
their outcomes is generally a finer measurement: it can reveal individual
data values and disturb previously imposed stabilizers.  We instead derive
a coherence-preserving construction from the ideal instrument
\eqref{eq:instrument_pol}.

First, local Hadamard rotations convert the logical \(XX\) primitive
\eqref{eq:parity_restricts_to_XX} into a \(ZZ\) measurement.  For a data
control \(q\), target \(r\), and ancilla \(a\) initialized in
\(|+_L\rangle_a\), perform, in this order,
\begin{equation}
 Z_L^{(q)}Z_L^{(a)}\ \text{with outcome }(-1)^u,\qquad
 X_L^{(a)}X_L^{(r)}\ \text{with outcome }(-1)^v,\qquad
 Z_L^{(a)}\ \text{with outcome }(-1)^w.
 \label{eq:measurement-CNOT-sequence}
\end{equation}
Define \(\Pi_b^P=(I+(-1)^bP)/2\).  Contracting the ancilla gives the
conditional Kraus operator on the data qubits,
\begin{equation}
 \begin{aligned}
 K_{uvw}
 &={}_a\langle w_L|\Pi_v^{X_aX_r}\Pi_u^{Z_qZ_a}|+_L\rangle_a\\
 &=\frac{(-1)^{v(u\oplus w)}}{2\sqrt2},
 (Z_L^{(q)})^v(X_L^{(r)})^{u\oplus w}\,
 \operatorname{CNOT}_L^{q\to r}.
 \end{aligned}
 \label{eq:measurement-CNOT-Kraus}
\end{equation}
This identity follows by applying both sides to the four computational
basis states.  Each outcome has probability \(1/8\), independently of the
input.  The known Pauli product can be corrected or tracked; hence the
sequence implements a logical CNOT using only two-body parity measurements,
a single-qubit ancilla preparation, and single-qubit readout and corrections.
This derivation does not assume the Choi resource constructed below.

To measure \(Z_L^{(q_1)}Z_L^{(q_2)}Z_L^{(q_3)}\), initialize a fresh
probe \(b\) in \(|0_L\rangle_b\), apply the corrected CNOT primitive from
each data qubit to the probe, and measure \(Z_L^{(b)}\).  With
\(U_{\mathrm{acc}}:=\prod_{j=1}^3\operatorname{CNOT}_L^{q_j\to b}\),
the resulting data Kraus operator is
\begin{equation}
 {}_b\langle z_L|U_{\mathrm{acc}}|0_L\rangle_b
 =\frac{I+(-1)^z Z_L^{(q_1)}Z_L^{(q_2)}Z_L^{(q_3)}}{2}.
 \label{eq:three-body-parity-instrument}
\end{equation}
Thus only the total parity is read out, and coherence inside its eigenspaces
is preserved.  Intermediate Pauli corrections from
\eqref{eq:measurement-CNOT-Kraus} are applied or propagated through the
remaining operations.

For \(\widetilde S_3\), choose \((q_1,q_2,q_3)=(C,C',D)\) and conjugate
the full instrument by Hadamards on \(C,C'\).  For \(\widetilde S_4\), use
\((D,D',C)\) and Hadamards on \(D,D'\).  Each construction gives exactly
the L\"uders instrument of the corresponding stabilizer.  Sequential
measurements therefore preserve the constraints of the other commuting
stabilizers.  After all four measurements, outcome signs can be changed to
\(+1\) by known local Pauli products, preparing the resource deterministically
in the ideal-instrument model.

\subsubsection{Parity Polarization Measurements and Controlled-$Z_L$ Gate}
	Each $\widetilde S_i$ is Hermitian with eigenvalues $\pm1$, and one checks directly that
	\(
	[\widetilde S_i,\widetilde S_j]=0
	\)
	for all $i,j$. Therefore the four observables admit a simultaneous eigenbasis. The eigenvector is unique because the four commuting stabilisers are
	independent and generate a stabiliser group of size $2^4$ acting on $4$ qubits, hence they
	fix a one--dimensional subspace in $\mathcal C^{\otimes 4}$. In order to find the form of $|R_{CZ}\rangle$ in logical basis note that:

	\[
	\widetilde S_1=+1\ \Rightarrow\ Z_C=Z_{C'},\qquad
	\widetilde S_2=+1\ \Rightarrow\ Z_D=Z_{D'}.
	\]
	
	Thus the support consists of the four basis states with equal bits within
each pair of registers $(C,C')$ and $(D,D')$.
	Write
	\begin{equation*}
		|R_{CZ}\rangle
		=
		a\,|00\rangle_{CC'}|00\rangle_{DD'}
		+b\,|00\rangle_{CC'}|11\rangle_{DD'}
		+c\,|11\rangle_{CC'}|00\rangle_{DD'}
		+d\,|11\rangle_{CC'}|11\rangle_{DD'}.
	\end{equation*}
	
	Now impose $\widetilde S_3=+1$ and $\widetilde S_4=+1$.
	Using that $X_CX_{C'}$ swaps $|00\rangle_{CC'}\leftrightarrow|11\rangle_{CC'}$ and that $Z_D$
	acts as $+1$ on $|00\rangle_{DD'}$ and $-1$ on $|11\rangle_{DD'}$, the constraint
	$\widetilde S_3|R_{CZ}\rangle=|R_{CZ}\rangle$ gives
	\begin{equation*}
		c=a,\qquad d=-b.
	\end{equation*}
	Similarly, since $X_DX_{D'}$ swaps $|00\rangle_{DD'}\leftrightarrow|11\rangle_{DD'}$ and $Z_C$
	acts as $+1$ on $|00\rangle_{CC'}$ and $-1$ on $|11\rangle_{CC'}$, the constraint
	$\widetilde S_4|R_{CZ}\rangle=|R_{CZ}\rangle$ gives
	\begin{equation*}
		b=a,\qquad d=-c.
	\end{equation*}
	Combining these constraints yields
	\(
	b=c=a
	\)
	and
	\(
	d=-a.
	\)
	After normalisation we obtain the explicit expression of $|R_{CZ}\rangle_{CC'DD'}$
	\begin{equation}
		|R_{CZ}\rangle_{CC'DD'}
		=
		\frac12\Big(
		|00\rangle_{CC'}|00\rangle_{DD'}
		+
		|00\rangle_{CC'}|11\rangle_{DD'}
		+
		|11\rangle_{CC'}|00\rangle_{DD'}
		-
		|11\rangle_{CC'}|11\rangle_{DD'}
		\Big).
		\label{eq:RCZ_explicit}
	\end{equation}
	
	We now show that \eqref{eq:RCZ_explicit} coincides with the Choi--Jamio\l kowski state of the
	unitary Controlled $Z_L$ quantum gate with input registers $(C',D')$ and output registers $(C,D)$.
	
	The action of a unitary $CZ_L$ gate on the logical basis two qubits basis is:
	\begin{equation*}
		CZ_L |x_L\rangle |y_L\rangle = (-1)^{xy}|x_L\rangle |y_L\rangle
	\end{equation*}
	
	Define the logical EPR pairs
	\begin{equation*}
		|\Phi^+\rangle_{CC'}=\frac{|00\rangle_{CC'}+|11\rangle_{CC'}}{\sqrt2},
		\qquad
		|\Phi^+\rangle_{DD'}=\frac{|00\rangle_{DD'}+|11\rangle_{DD'}}{\sqrt2}.
	\end{equation*}
	Applying \(CZ_L\) to the output registers multiplies a computational
basis term by \((-1)^{cd}\).  Comparing with \eqref{eq:RCZ_explicit} gives
\begin{equation}
 |R_{CZ}\rangle_{CC'DD'}=
 CZ_L^{(C,D)}\bigl(|\Phi^+\rangle_{CC'}\otimes|\Phi^+\rangle_{DD'}\bigr),
 \label{eq:R_CZ_def}
\end{equation}
where the gate acts as the identity on \(C',D'\).  This is the normalized
Choi state of \(CZ_L\); register superscripts specify the tensor ordering.

\subsubsection{Controlled NOT Gate via Teleportation}
	
	Exploiting the above construction we are now going to define a teleportation protocol that implements a $CNOT_L$ on an arbitrary two qubits state $|\psi\rangle_{AB}$ teleporting it on two output qubits state $|\text{out}\rangle_{CD}$.

	Let $(A,B)$ be the two \emph{input} logical SPE qubits in an arbitrary state $|\psi\rangle_{AB}$.
	Assume we have prepared the four--qubit resource $|R_{CZ}\rangle_{CC'DD'}$.
	
	We now perform logical Bell measurements on the pairs $(A,C')$ and $(B,D')$.
	A logical Bell measurement on a pair $(q,r)$ is equivalent to measuring the commuting observables
	\begin{equation*}
		Z_L^{(q)}Z_L^{(r)}\quad\text{and}\quad X_L^{(q)}X_L^{(r)}.
	\end{equation*}
	In our architecture, $X_LX_L$ is directly measured by polarisation parity
	(see ~\eqref{eq:parity_restricts_to_XX}), while $Z_LZ_L$ is obtained by conjugation with
	$H_L\otimes H_L$ (available from single--qubit $SU(2)$ control).
	
	Encode the Bell outcomes on $(A,C')$ as bits $(x,y)\in\{0,1\}^2$:
	\begin{equation}
		Z_L^{(A)}Z_L^{(C')}=(-1)^x,
		\qquad
		X_L^{(A)}X_L^{(C')}=(-1)^y,
		\label{eq:Bell_bits_ACp}
	\end{equation}
	and similarly on $(B,D')$ as $(l,m)\in\{0,1\}^2$:
	\begin{equation}
		Z_L^{(B)}Z_L^{(D')}=(-1)^l,
		\qquad
		X_L^{(B)}X_L^{(D')}=(-1)^m.
		\label{eq:Bell_bits_BDp}
	\end{equation}
	
	After these Bell measurements, the measured qubits are $A,C',B,D'$, hence they are no longer part
	of the quantum output. The only remaining quantum system is precisely the pair $(C,D)$.
	
	With the outcome convention above, each Bell contraction contributes
\(X^xZ^y\) or \(X^lZ^m\) before the gate.  Commuting these through
\(CZ_L\) propagates each \(X\) into an additional \(Z\) on the other
output.  Up to a global phase, the normalized conditional output is
	\begin{equation}
 |\mathrm{out}\rangle_{CD}=
 \bigl(X_L^x Z_L^{y\oplus l}\bigr)^{(C)}
 \bigl(X_L^l Z_L^{m\oplus x}\bigr)^{(D)}
 CZ_L|\psi\rangle_{AB},
 \label{eq:teleported_CZ_final}
\end{equation}
	Here $CZ_L|\psi\rangle_{AB}$ is understood as the same logical vector, with the information
	teleported from registers $(A,B)$ onto $(C,D)$, and acted on by $CZ_L$.
	This equation implements $CZ_L$ up to Pauli byproducts: the byproducts are
	known local Pauli operators determined by the classical outcome bits $(x,y,l,m)$, and are
	therefore exactly correctable by a final single--qubit correction or by tracking a classical Pauli frame.
	
	Finally,
	\begin{equation*}
		CNOT_L = \big(\mathbf 1\otimes H_L\big)\;
		CZ_L\;
		\big(\mathbf 1\otimes H_L\big),
	\end{equation*}
	so the parity--based construction yields a logical $CNOT_L$ gate on the SPE code.

	\subsection{Circuit-model universality}
	Since we can implement arbitrary single--qubit unitaries $SU(2)$ on $\mathcal C$
	(section ~\ref{subsec:blochsphere}) and we can implement an entangling two--qubit gate $CNOT_L$
	(section ~\ref{subsec:RCZ_stabiliser}), the SPE model is universal for quantum computation in the
	standard circuit sense.
	
	\section{Conclusion}
	
	We have studied the massless sector of the
\(C_{\hat n}\)-extended Poincar\'e group
\eqref{eq:gqc-Cn-extended-poincare-group}.  Clifford--Mackey theory pairs
the future- and past-directed Wigner representations into a single
extended UIR.  The eigenvalues \(\varepsilon=\pm1\) of the corresponding
exchange operator distinguish states within that representation, while
the point little group remains \(ISO(2)\).  Independently accessible
observable algebras select the tensor-product structure and establish
when these exchange states are entangled.

The finite-mode map \eqref{eq:finite-mode-realization-map} realizes the
selected exchange and observable structure with positive-frequency
propagation modes and polarization.  The joint correlation
\eqref{eq:XX-expectation} distinguishes the two selected optical eigenstates.
This is an accessible optical encoding, not an identification of a
past-directed Wigner momentum with a positive-energy photon, nor a standalone
experimental discrimination of the spacetime extension.

As emphasized in Section~\ref{sec:spe-one-qubit-distinction}, this is a
one-logical-qubit-per-photon encoding: the two logical basis values are
selected by the positive or negative $X$-basis path--polarization
correlation, not by an outcome of either factor separately.

The logical code \eqref{eq:logical-code} supports the full Bloch sphere
under the specified coherent controls.  It is a selected two-dimensional
subspace of the four-dimensional optical carrier, not the entire
infinite-dimensional Poincar\'e representation space.  Ideal controls
preserve it by construction; this alone does not establish suppression of
physical leakage or a scalability advantage.

Under the explicit assumption of a nondestructive, coherence-preserving
parity instrument, \eqref{eq:measurement-CNOT-Kraus} constructs an entangling
logical gate and \eqref{eq:three-body-parity-instrument} supplies the required
stabilizer measurements.  The Choi resource \eqref{eq:R_CZ_def} then gives
the teleportation protocol \eqref{eq:teleported_CZ_final}, with outcome-dependent
Pauli corrections.  Together with arbitrary single-qubit control, these
operations establish circuit-model universality.  The resource cost, success
probability and noise performance of a physical parity device remain
experimental questions beyond this ideal computational construction.

\clearpage

	%---------------------------------------------------------

	\end{document}